\documentclass[RNAAS]{aastex63}
\usepackage{amsmath,amssymb,CJK}

\begin{document}
\begin{CJK*}{UTF8}{gbsn}

\title{Recovery of Returning Halley-Type Comet 12P/Pons-Brooks With the Lowell Discovery Telescope}

\correspondingauthor{Quanzhi Ye}
\email{qye@umd.edu}

\author[0000-0002-4838-7676]{Quanzhi Ye (叶泉志)}
\affiliation{Department of Astronomy, University of Maryland, College Park, MD 20742, USA}

\author[0000-0002-4767-9861]{Tony L. Farnham}
\affiliation{Department of Astronomy, University of Maryland, College Park, MD 20742, USA}

\author[0000-0003-2781-6897]{Matthew M. Knight}
\affiliation{Department of Physics, United States Naval Academy, 572C Holloway Rd, Annapolis, MD 21402, USA}
\affiliation{Department of Astronomy, University of Maryland, College Park, MD 20742, USA}

\author{Carrie E. Holt}
\affiliation{Department of Astronomy, University of Maryland, College Park, MD 20742, USA}

\author{Lori M. Feaga}
\affiliation{Department of Astronomy, University of Maryland, College Park, MD 20742, USA}



\begin{abstract}
    We report the recovery of returning Halley-type comet 12P/Pons-Brooks using the 4.3~m Lowell Discovery Telescope, at a heliocentric distance of 11.89~au. Comparative analysis with a dust model suggests that the comet may have been active since $\sim30$~au from the Sun. We derive a nucleus radius of $17\pm6$~km from the nucleus photometry, though this number is likely an overestimation due to the contamination from dust and gas. Continuing monitoring is encouraged in anticipation of the comet's forthcoming perihelion in 2024 April.
\end{abstract}


\section{}

Halley-type comet 12P/Pons-Brooks (hereafter 12P) has been linked to observations dating back to 1385 A.D. and possibly, to observations in 245 A.D. \citep{Green2020a, Nakano2020}\footnote{See also \url{http://www.comethunter.de/12P.pdf}, accessed 2020 June 30. Details will be reported in M. Meyer et al. (in preparation).}, making it the comet with the second longest observational arc of all known comets, after only 1P/Halley\footnote{If we discount the linkage to the 245 A.D. observations, then 12P will fall short of 153P/Ikeya-Zhang \citep[with observations dating back to 877 A.D.,][]{Hasegawa2003} and be the comet with the third longest arc of all known comets.}. 12P was last observed in 1953--54 during its previous perihelion passage. Its next perihelion is in 2024 April of which it is predicted to reach $\sim4$~magnitude\footnote{\url{http://www.aerith.net/comet/catalog/0012P/2024.html}, accessed 2020 June 30.}.

We recovered 12P using the 4.3~m Lowell Discovery Telescope (LDT; formerly known as the Discovery Channel Telescope, DCT) on 2020 June 10 and 17 UTC. At the time of the observations, the comet was at $r_\mathrm{H}=11.89$~au from the Sun, slightly farther than the 1982 recovery of 1P/Halley's 1985--86 perihelion \citep[at 11.04~au;][]{Jewitt1982}, making it the most distant recovery of a returning periodic comet to-date of which we are aware. Observations were made using the Large Monolithic Imager \citep[LMI;][]{Massey2013} and the broadband {\it VR} filter. LMI has a field-of-view (FOV) of $12.3'\times12.3'$ and a pixel scale of $0.36''$ after an on-chip $3\times3$ binning. Images were bias-subtracted and flat-field corrected using {\tt ccdproc} \citep{Craig2015}, and were then calibrated using the {\tt PHOTOMETRYPIPELINE} code \citep{Mommert2017} with the Gaia-DR2 catalog \citep{Gaia2018} for astrometry and the {\it r}-band PanSTARRS DR1 catalog \citep{Magnier2016} for photometry, respectively. For photometric calibration, we used only stars with Sun-like colors ($g_\mathrm{PS1}-r_\mathrm{PS1} \in (0.2, 0.6)$ whereas Sun's $g_\mathrm{PS1}-r_\mathrm{PS1}=0.4$) in order to mitigate the potential color systematic bias due to the approximation between {\it r} and the non-standard {\it VR}. Astrometry and improved orbit is published in Minor Planet Electronic Circular 2020-M114\footnote{\url{https://www.minorplanetcenter.net/mpec/K20/K20MB4.html}, accessed 2020 June 30.} and \citet{Green2020b}.

12P was found $\sim8''$ ($\sim15\sigma$) south of the nominal position predicted by JPL Orbit Solution \#15 (with data arc from 1812 to 1954). This relatively large offset is likely caused by multiple factors arising from previous astrometric measurements, such as the difficulty to accurately measure the position of cometary nucleus, lower-precision astrometric reference catalogs used for previous apparitions, and non-gravitational forces acting on the nucleus, which have been thoroughly discussed in previous works \citep[e.g.][]{Farnocchia2016, Li2017}. Our experience with 12P adds to this consensus that it is important to reserve some margin in order to account for these effects in recovery works, especially for comets that lack modern CCD observations.

\begin{figure}[h!]
\begin{center}
\includegraphics[scale=0.6,angle=0]{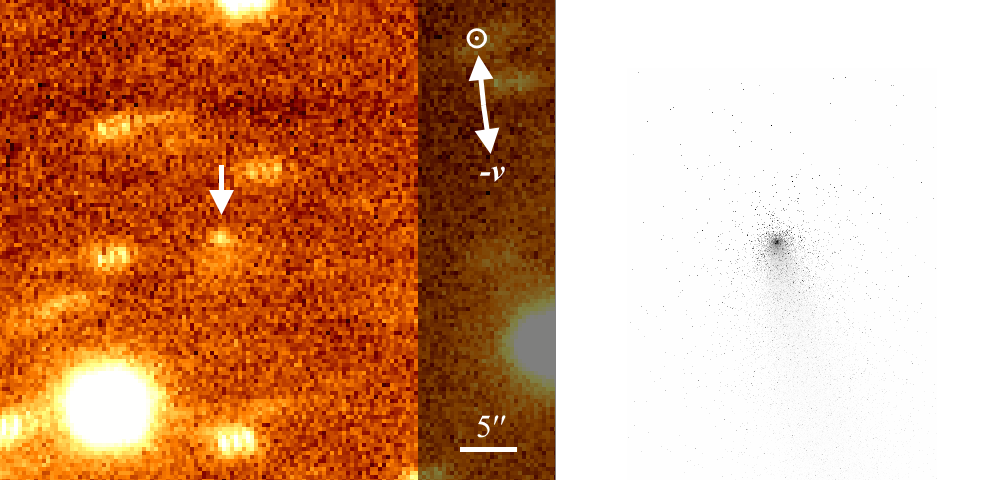}
\caption{Left panel: the composite LDT/LMI image of 12P on 2020 June 17, with the comet marked by an arrow, and the direction to the Sun and negative velocity vector marked by $\odot$ and $-v$. Right panel: a crude best-fit dust model at the same scale of the left panel. This image is made with five 300-second exposure with the {\it VR} filter. The best-fit dust model assumes a start of activity at $r_\mathrm{H}=30$~au and a minimum grain size of $10~\micron$, with a constant ejection speed of 1~m/s and a sunward ejection cone with an opening angle of $45^\circ$. The dust modeling is conducted using a modified {\tt MERCURY6} code \citep{Chambers1999, Ye2016}. \label{fig:img}}
\end{center}
\end{figure}

Our initial attempt on June 10 successfully located 12P, but the images were severely degraded by background sources and sub-optimal sky conditions. We then conducted a target-of-opportunity (ToO) session on June 17 under improved conditions and acquired additional images. Figure~\ref{fig:img} shows the aligned and median combined image using the June 17 data. The head of the comet has a full-width-half-maximum (FWHM) of $\sim1.2''$, largely comparable to the sizes of background stars, indicating the lack of a resolvable coma. A broad tail, measuring $\sim3''$ in length, is seen pointing towards $\sim170^\circ$--$200^\circ$ east of north, encompassing both the anti-sunward ($186^\circ$) and negative velocity vectors ($199^\circ$), implying a wide range of dust sizes and that the comet has been active for some time. A crude dust model, assuming CO as the dominant volatile species driving the activity, suggests that 12P may have been active since $r_\mathrm{H}\approx30$~au, though we note that this is highly dependent on the details of the mechanism driving the activity. With observations that are only one week apart, we also cannot distinguish between continuous activity and an outburst. Interestingly, 1P/Halley was also observed to be active at $r_\mathrm{H}=28.1$~au \citep{Hainaut2004} though this was an outburst detected in its outbound leg.

Using a $1.4''$-wide aperture, the nucleus-only brightness of 12P on the 2020 June 17.32 UTC is $r_\mathrm{PS1}=24.16\pm0.03$. The cross-section area $C$ can be calculated from

\begin{equation}
    C = \frac{\pi r_\mathrm{H}^2 \varDelta^2}{p_\lambda (1~\mathrm{au})^2 \phi(\alpha)} 10^{-0.4(m_r-m_{\odot, r})}
\end{equation}

\noindent where $r_\mathrm{H}=11.89$~au and $\varDelta=11.15$~au are the heliocentric and geocentric distances of 12P, respectively, $p_\lambda=0.04$ is the typical geometric albedo of cometary nuclei, $\phi(\alpha)=0.035 \alpha~\mathrm{mag/deg}$ is a simple phase function with $\alpha=3.4^\circ$ being the phase angle, $m_r$ is the brightness of the nucleus, and $m_{\odot, r}=-26.93$ is the $r$-band magnitude of the Sun \citep{Willmer2018}. By inserting all numbers, we derive $C=950\pm630~\mathrm{km^2}$, which can be translated to a nucleus with radius $R_\mathrm{N}=17\pm6~\mathrm{km}$. This is about two times larger than 1P/Halley \citep[$16\times8\times8$~km in extent;][]{Keller1987}, though we note that this number is perhaps best regarded as an upper limit due to the undercorrected contamination from cometary dust and gas.

12P has reportedly experienced multiple outbursts during its 1883--84 and 1953--54 perihelion passages \citep{Whitney1955, Kronk2003, Kronk2009}. Continuing monitoring is encouraged as we are approaching its next perihelion in 2024.

\acknowledgments

We thank Maik Meyer and Michael Kelley for helpful discussion and inputs, as well as LaLaina Shumar, Jason Sanborn and Ana Hayslip for operating the telescope. We are grateful to Jennifer Hanley for kindly accommodating our target-of-opportunity observation. The University of Maryland observing team consisted of Lori Feaga, Quanzhi Ye, James Bauer, Tony Farnham, Carrie Holt, Michael Kelley, Jessica Sunshine, and Matthew Knight. These results made use of the Lowell Discovery Telescope (LDT) at Lowell Observatory. Lowell is a private, non-profit institution dedicated to astrophysical research and public appreciation of astronomy and operates the LDT in partnership with Boston University, the University of Maryland, the University of Toledo, Northern Arizona University and Yale University. The Large Monolithic Imager was built by Lowell Observatory using funds provided by the National Science Foundation (AST-1005313). This research has made use of data and services provided by the International Astronomical Union's Minor Planet Center (\url{https://minorplanetcenter.net/data}).

\software{{\tt ccdproc} \citep{Craig2015}, {\tt MERCURY6} \citep{Chambers1999}, {\tt PHOTOMETRYPIPELINE} \citep{Mommert2017}}
\facilities{LDT}

\bibliographystyle{aasjournal}
\bibliography{rnaas}{}

\end{CJK*}
\end{document}